# Measurement Scale Effect on Prediction of Soil Water Retention Curve and Saturated Hydraulic Conductivity

Behzad Ghanbarian[1*], Vahid Taslimitehrani[2], Guozhu Dong[2], Yakov A. Pachepsky[3]

[1] Dept. Petroleum & Geosystems Eng., University of Texas at Austin TX 78712

[2] Dept. Computer Sci. & Eng., Wright State University, Dayton OH 45435

[3] Environmental Microbial Safety Lab., USDA-ARS, Beltsville, MD 20705

[*] Corresponding author's email address: ghanbarian@austin.utexas.edu

**Abstract**

Soil water retention curve (SWRC) and saturated hydraulic conductivity (SHC) are key hydraulic properties for unsaturated zone hydrology and groundwater. In particular, SWRC provides useful information on entry pore-size distribution, and SHC is required for flow and transport modeling in the hydrologic cycle. Not only the SWRC and SHC measurements are time-consuming, but also scale dependent. This means as soil column volume increases, variability of the SWRC and SHC decreases. Although prediction of the SWRC and SHC from available parameters, such as textural data, organic matter, and bulk density have been under investigation for decades, up to now no research has focused on the effect of measurement scale on the soil hydraulic properties pedotransfer functions development. In the literature, several data mining approaches have been applied, such as multiple linear regression, artificial neural networks, group method of data handling. However, in this study we develop pedotransfer functions using a novel approach called contrast pattern aided regression (CPXR) and compare it with the multiple linear regression method. For this purpose, two databases including 210 and 213 soil samples are collected to develop and evaluate pedotransfer functions for the SWRC and SHC, respectively, from the UNSODA database. The 10-fold cross-validation method is applied to evaluate the accuracy and reliability of the proposed regression-based models. Our results show that including measurement scale parameters, such as sample internal diameter and length could substantially improve the accuracy of the SWRC and SHC pedotransfer functions developed using the CPXR method, while this is not the case when MLR is used. Moreover, the CPXR method yields remarkably more accurate soil water retention curve and saturated hydraulic conductivity predictions than the MLR approach.

## 1. Introduction

Flow and transport modeling in saturated and unsaturated environmental hydrologic large watershed scales requires characterization of hydraulic properties at smaller pore and core scales. One of these hydraulic properties is soil water retention curve (SWRC) whose measurement, estimation, and even modeling are still under consideration in different communities, such as hydrology, soil science, and hydrogeology. In order to measure the typical drying branch of SWRC, either water can be drained from the sample, or mercury can be injected into it. However, the pore-size distribution deduced from the soil water retention curve may be different from the real pore size distribution considerably (Dullien, 1975) due to complicated phenomena, such as trapping and wettability. As Dullien (1975) stated, the soil water retention curve yields the entry pore-size distribution rather than the true pore-size distribution. Hall et al. (1986) presented results indicating that pore-size distributions measured by SANS and SAXS techniques were in reasonable agreement with those measured by nitrogen adsorption isotherm. However, those distributions measured by nitrogen desorption isotherm and mercury porosimetry were different.

Sample thickness (measurement scale) also influences soil water retention curve measurements. Larson and Morrow (1981) demonstrated that the soil water retention curve became sharper at high saturations as the measurement scale decreased (see Larson and Morrow (1981) and Hunt et al. (2013a) for further discussion). This finite-size (or measurements scale) effect on air-entry value might be one of the reasons that existing parametric pedotransfer functions are inaccurate in estimation of air-entry value from other easily available parameters (see e.g., Fredlund et al., 2002; Ghanbarian-Alavijeh et al., 2010).

Issues with the soil water retention curve measurements are not restricted to the wet end. Recently, Bittelli and Flury (2009) reported remarkable differences between pressure plate and dew point meter measurements especially at the dry end (e.g., when tension head is greater than 100 kPa). They found that water content determined using pressure plate at suction 1500 kPa was two times larger than that measured with dew point meter. This might be due to lack of equilibrium at relatively high tensions. Considering all these issues with the soil water retention curve (SWRC), this hydraulic characteristic still provides useful information e.g., entry pore-size distribution, which can be applied to modeling flow and transport in porous media e.g., unsaturated hydraulic conductivity prediction (Burdine, 1953; Mualem, 1976; van Genuchten, 1980; Hunt, 2001; Assouline and Tartakovsky, 2001; Ghanbarian-Alavijeh and Hunt, 2012; Hunt et al., 2013b).

In addition to SWRC, saturated hydraulic conductivity (SHC) plays a key role in flow and solute transport modeling for under saturated and unsaturated conditions. In particular, unsaturated hydraulic conductivity predictions require the SHC value, and contaminant transport e.g., dispersion modeling in groundwater does so.

The effect of measurement scale on the saturated hydraulic conductivity value attracted a great deal of attention. It has been shown that the SHC measurement is much influenced by sample size. Mallants et al. (1997) investigated the spatial variability of the SHC measurement using columns of different lengths and diameters, e.g., 5.1×5 cm, 20×20 cm, and 100×30 cm. They found that the geometric mean of SHC decreased as the column size increased. In another study, Lai and Ren (2007) reported similar results using double-ring infiltrometers in the field scale. They found that in highly heterogeneous soils infiltrometers should have inner ring diameter larger than 80 cm to obtain reliable measurements.

Hillel (2004) pointed out that hydraulic conductivity measurement scale dependence is due to inhomogeneity. He explains, "owing to soil heterogeneity, the apparent hydraulic conductivity measured often depends on the scale of the measurement. Thus, the $K$ [hydraulic conductivity] value measured on a cubic centimeter or decimeter may differ from the average value measured on a cubic meter. Too often, this is ignored and $K$ values are reported without specifying the scale of the measurement." Recently, Hunt (2006) proposed a scale-dependent hydraulic conductivity model for anisotropic media. His percolation-based model compared reasonably well with experiments.

Since both soil water retention curve (SWRC) and saturated hydraulic conductivity (SHC) measurements are time consuming, indirect methods have been developed to estimate SWRC and SHC from other available properties, such as sand, silt, and clay contents, organic matter, bulk density, and particle-size distribution. For this purpose, various statistical and data mining approaches have been applied in the literature, such as multiple linear regression (Wösten et al., 1999; Vereecken and Herbst, 2004; Ghanbarian-Alavijeh and Millán, 2010), neural networks models (Pachepsky et al., 1996; Schaap and Bouten, 1996; Schaap and Leij, 1998; Parasuraman et al., 2006; Ghanbarian-Alavijeh et al., 2012), group method of data handling (Pachepsky et al., 1998; Pachepsky and Rawls, 1999; Ungaro et al., 2005), nonparametric $k$-nearest neighbor approach (Nemes et al., 2006; Botula et al., 2013), and supporting vector machines (Lamorski et al., 2008; Twarakavi et al., 2009).

Pachepsky et al. (2001) state, "… observations indicate the need to pay more attention to the effects of scale on soil hydraulic properties and the need to include such effects in pedotransfer functions." However, to our knowledge up to now no research has focused on the effect of measurement scale upon the development of the soil hydraulic properties pedotransfer functions. Therefore, the main objectives of this study are to: (1) propose measurement-scale dependent pedotransfer functions for predicting the soil water retention curve (SWRC) and the saturated hydraulic conductivity (SHC), and (2) introduce and evaluate a new data mining approach called contrast pattern aided regression (CPXR) for soil hydraulic properties predictions using a large number of data available in the UNSODA database.

**2. Contrast pattern aided regression approach**

The contrast pattern aided regression (CPXR) method, introduced by Dong and Taslimitehrani (2014), is a novel, robust, and powerful regression-based method for building prediction models. CPXR has several significant advantages including high prediction accuracy and ability to deal with highly complicated and diverse predictor-response relationships, and less over-fitting. The prediction models returned by CPXR are also representable, in contrast to artificial neural network models. Extensive experiments demonstrated that (Dong and Taslimitehrani, 2014) the CPXR-based models are more accurate than those developed using other methods, such as piecewise linear regression, support vector regression, gradient boosting, and Bayesian additive regression trees. The CPXR method has been also applied in other fields and showed better performance comparing to other classifiers (see e.g., Taslimitehrani and Dong, 2014). The main idea of CPXR is to use a pattern, conjunction of several conditions on a small number of predictor variables, as logical characterization of a subgroup of data, and a local regression model (corresponded to pattern) as a behavioral characterization of the predictor-response relationship for data instances of that subgroup of data. What makes CPXR a powerful technique is it can pair a pattern and a local regression model to represent a specific predictor-response relationship for a subgroup of data. It also has the flexibility in pairing multiple patterns and local regression models to represent distinct predictor-response relationships for multiple subgroups of data. Another reason why

CPXR substantially outperforms most of other prediction methods is that it uses an effective mechanism to select a highly collaborative set of a small number of patterns to maximize their overall combined prediction accuracy. In the following section, we introduce the CPXR technique. However, for more technical detail, the interested reader is referred to the Dong and Taslimitehrani (2014) article.

**2.1. CPXR Algorithm**

Assume a database $D$ consisting of $N$ pairs ($x_n$, $y_n$) in which $y_n$ is a vector including desirable outputs and $x_n$ is a vector of input parameters. We define *item* as a conditional input parameter of the form of "$A = a$", if $A$ is a categorical parameter, or "$v_1 \leq A < v_2$", if $A$ is a numerical parameter, where $a$, $v_1$, $v_2$ are constants. A *pattern* is therefore a finite set of items. In what follows, we provide an example illustrating how item and pattern are defined. Consider a dataset (see Table 1), which consists of four soil samples with three measured parameters: sand, silt, and clay. "82 < Sand < 86" and "Clay = 3" are example of a numerical and a categorical parameter, respectively, and "82 < Sand < 86 AND Clay = 3" is an example of pattern as a conjunction of items. A sample would *match* a pattern, if every condition in such a pattern is true. For example, sample #1 belongs to the corresponding pattern defined above (82 < Sand < 86 AND Clay = 3). Sample #4, however, does not because although its sand content ranges between 82 and 86, the clay content (Clay = 1) does not match the second item (Clay = 3). The *matching dataset* (mds($p$, $D$)) of pattern $p$ in database $D$ is, therefore, a set of all samples matched in pattern $p$.

**2.2. PXR models**

The main idea of the CPXR method is developing local regression models for all database subgroups, which are the matching dataset of the recognized patterns, then using patterns to logically characterize each subgroup. The following definition shows that a (local) multiple linear regression model corresponding to each pattern $p$ characterizes the behavior of samples matched in pattern $p$.

Given a training database $D_t$ for the regression purpose, a pattern aided regression (PXR) model is represented by a tuple PM = $((p_i, f_i, w_i),..., (p_k, f_k, w_k), f_d)$ in which $k$ is the number of recognized patterns, $p_i$ is the pattern, $f_i$ is the local multiple linear regression model, and $w_i > 0$ is the weighting factor. $f_d$ is the default model, in other words the regression model developed to characterize those samples, which do not match any pattern. The regression function of PM is given (for each sample $X$) by

$$f(X) = \begin{cases} \dfrac{\sum\limits_{p_i \in \pi_X} w_i f_i(X)}{\sum\limits_{p_i \in \pi_X} w_i}, & \pi_X \neq \emptyset \\ f_d(X), & \text{otherwise} \end{cases} \quad (1)$$

where $\pi_X = \{1 \leq i \leq k,\ X \text{ satisfies } p_i\}$.

Equation (1) means if a sample matches just one pattern, to predict the desirable output the corresponding multiple linear function $f_i$ is applied. However, when a sample matches more than one pattern, weighted averaging is used to determine desirable output. If a sample does not match any pattern, then the default $f_d$ is used.

Dong and Taslimitehrani (2014) stated that highly complex databases contain multiple subgroups whose best-fit local multiple linear regression models behave substantially different. Those behaviors are called diverse predictor-response relationships. Each pattern and its local model pair in a PXR model are intended to represent the predictor-response relationship for one subgroup of data (i.e., the pattern's matching dataset).

**2.3. Sketch of CPXR**

In the following, we present 5 steps to show how the CPXR algorithm is designed to build a PXR model by computing an optimal pattern set, which defines a PXR model associated with minimal total errors.

1- CPXR starts with a *baseline regression model* $f_0$ built on the training database $D_t$ using multiple linear regression technique. Then, samples are sorted based on the errors of the corresponding baseline regression model, and cumulative error is calculated.

2- An arbitrary value of 45 percent of the cumulative error is chosen to find the cutting point, which divides the training database Dt into two classes: LE (large errors) and SE (small errors). LE contains those samples whose cumulative error is greater than 45 percent of the cumulative error. Note that 45% is an optimized value found by analyzing more than 50 different databases in various research fields (see Dong and Taslimitehrani, 2014).

3- In order to discretize input variables and define items, an entropy-based binning method (Fayyad and Irani, 1993) is used. Then, CPXR mines all contrast patterns of the LE class (Li et al., 2005). Since those patterns are more frequent in LE than in SE, they are likely to capture subgroups of data where *f0* makes large prediction errors. Several filters are also used to remove those patterns, which are very similar to others.

4- Then a local multiple linear regression model $f_i$ is built for each remaining contrast pattern $p_i$. Those patterns and local multiple linear regression models, which do not improve the accuracy of predictions, are removed at this step.

5- CPXR then applies a double (nested) loop to search for an optimal pattern. For this purpose, we replace a pattern by another one in the pattern set to minimize errors in each iteration.

## 3. Materials and Methods

In this study, we selected those experiments from the UNSODA database (Leij et al., 1996; Nemes et al., 2001) whose measurement scale data were available. Two databases including 210 and 213 samples were collected to develop and evaluate pedotransfer functions for the soil water retention curve (SWRC) and saturated hydraulic conductivity (SHC), respectively. The common method of 10-fold cross-validation (Han et al., 2011) was applied to evaluate the accuracy and reliability of the developed pedotransfer functions. For this purpose, each original dataset was randomly divided into 10 equal sized subsamples: 8 subsamples used to train the model and 2 subsamples to test it. To minimize the effect of overfitting, this process was repeated 10 times, and statistical parameters such as root mean square error (RMSE), root mean square logarithmic error (RMSLE), and correlation coefficient ($R^2$) were determined each time.

The van Genuchten soil water retention curve model (van Genuchten, 1980; hereafter vG) parameters, such as $α$, $n$, $θ_r$, and $θ_s$ were obtained by directly fitting the vG model to the measured soil water retention curve data. These optimized parameters were also used to calculate water contents at different tension heads e.g., inflection point, 10, 33, 50, 100, 300, 500, 1000, 1500 kPa for each soil sample (Wösten and Nemes, 2004).

To develop both point and parametric pedotransfer functions for SWRC, 6 input variables, such as sand, silt, and clay contents, geometric mean diameter, geometric standard deviation, and bulk density were used (hereafter SWRC1). Geometric mean particle-size diameter, $d_g$ (mm), and geometric standard deviation, $\sigma_g$ (mm), were determined from clay, silt, and sand contents (Shirazi and Boersma, 1984). In order to investigate the effect of measurement scale on the development and evaluation of the pedotransfer functions, in addition to those 6 variables, sample internal diameter (ID) and length ($L$) were also included as input variables (hereafter SWRC2). In addition to point pedotransfer functions, we also developed parametric pedotransfer functions to predict the van Genuchten soil water retention curve model parameters, such as $\theta_r$, $\theta_s$, $\alpha$, and $n$ from available soil properties. The same input variables introduced to SWRC1 and SWRC2 models were used to develop two parametric models (hereafter SWRC3 and SWRC4). The difference between the SWRC3 and SWRC4 models is that two more input variables e.g., sample internal diameter (ID) and length ($L$) were used in in the development of the SWRC4 model.

For the development of the SHC pedotransfer functions four models were considered: SHC1 included input variables, such as sand, silt, and clay contents, geometric mean diameter, geometric standard deviation, and bulk density. SHC2 consisted of two extra input variables e.g., sample length and internal diameter. In SHC3, besides textural data e.g., sand, silt, and clay contents, geometric mean diameter, geometric standard deviation, and bulk density, vG soil water retention curve model parameters e.g., $\alpha$, $n$, $\theta_r$, and $\theta_s$ were also used. SHC4 included sample diameter and length in addition to all input variables applied in SHC3. To develop pedotransfer functions for saturated hydraulic conductivity, the logarithmic measured SHC values were used in the train and test processes. Table 2 summarizes the input and output variables for all models developed in this study.

In order to evaluate the performance of the contrast pattern aided regression (CPXR) technique, we also developed models described above e.g., SWRC1, SWRC2, SWRC4, SHC1, SHC2, SHC3, and SHC4 using the multiple linear regression (MLR) method. Statistical parameters, such as root mean square error (RMSE), root mean square logarithmic error (RMSLE), and correlation coefficient (R2) were calculated to compare the CPXR approach with the MLR technique.

## 4. Results and Discussion

Strictly speaking, the obtained results indicate that the prediction of soil water retention curve and saturated hydraulic conductivity improved remarkably when measurement scale parameters, such as sample internal diameter (ID) and length ($L$) are included in the pedotransfer functions development. We also find that CPXR is a robust, efficient approach to detect patterns in complex soil systems. Accuracy and reliability of the pedotransfer functions developed for both soil water retention curve and saturated hydraulic conductivity demonstrate that the CPXR technique is superior to the MLR method. The details of results are presented in Tables 3 to 6 for point and parametric pedotransfer functions developed for the soil water retention curve, and those developed for the saturated hydraulic conductivity. Note that the root mean square error (RMSE; for soil water retention), root mean square logarithmic error (RMSLE; for saturated hydraulic conductivity), and $R^2$ (correlation coefficient) values reported in Tables 3 to 6 are the average over 100 iterations (i.e., 10 repetitions of 10-fold cross-validations). In what follows, we discuss the obtained results in more detail.

**4.1. Point pedotransfer functions for soil water retention curve (SWRC)**

We found that inclusion of sample internal diameter (ID) and length ($L$) as input variables increased considerably the accuracy and reliability of the SWRC2 pedotransfer function compared to the SWRC1 one, in which ID and $L$ were not included, using CPXR. Although following Larson and Morrow (1981) one may expect including measurement scales affect the shape of soil water retention curve near saturation, we found such an influence at both low and high tension heads (see Table 3). For example, at $\theta_s$ and $\theta_i$ the RMSE values in the test (train) process decreased only by 11% (15%) and 15% (7%), respectively, while at $\theta_{10}$ and $\theta_{1500}$ the RMSE values decreased by 40% (31%) and 23% (32%) after we included measurement scale variables (i.e., ID and $L$). Comparison of $R^2$ values of SWRC1 and SWRC2 models (presented in Table 3) for the train and test processes also show that the accuracy of pedotransfer functions increased with including two sample internal diameter and length variables.

The graphical results of the SWRC1 and SWRC2 models developed using the CPXR method (the predicted water content as a function of the measured one at different tension heads) for *one iteration subset* are shown in Figs. 1a and 1b, respectively. The RMSE values presented in Figs. 1(a) and 1(b) are not different. However, as was demonstrated in Table 3 (results are average over 100 iterations; 10-fold cross-validation), sample internal diameter (ID) and length ($L$) play important role in the prediction of soil water content at different tension heads.

The SWRC1 model developed in this study is comparable to Model 1 proposed by Ghanbarian-Alavijeh and Millán (2010). Although Ghanbarian-Alavijeh and Millán (2010) used a larger database including 315 samples from the UNSODA database, both SWRC1 and Model 1 predict water content at the same tension heads using the same input variables. We compared the RMSE and $R^2$ values presented in Table 3 with those reported by Ghanbarian-Alavijeh and Millán (2010; see their Table 3). Our comparison shows that the CPXR method was trained and tested more accurately than the MLR method and stepwise technique used by Ghanbarian-Alavijeh and Millán (2010). Comparison of the RMSE values given in Table 3 with those reported by Lamorski et al. (2008) in their Table 1 indicates that the point pedotransfer functions developed by the CPXR method in this study predicted the water contents more accurately than their models developed by the artificial neural networks and supporting vector machines at all tension heads, except the saturation point.

Pachepsky et al. (2001) demonstrated that soil water retention curve measured at field scale might be different than that measured at lab scale because of different soil volumes and spatial scales. They compared soil water retention at field and lab scales and found the average difference between field and lab water contents near zero for coarse-textured soils. However, for fine-textured soils with sand content less than 50% field measurements were remarkably less than lab ones for water contents between 0.45 and 0.60 $cm^3$ $cm^{-3}$. Pachepsky et al. (2001) also found that a fractal bulk density-size scaling could describe the difference between field and lab measurements for the range of high water contents. They particularly emphasized on the effects of measurement scale on soil hydraulic properties and the need to include such effects in developing pedotransfer functions, demonstrated in this study for both soil water retention curve and saturated hydraulic conductivity.

Results of the MLR method used in this study for models SWRC1 and SWRC2 are presented in Table 4. In the best case (see $\theta_{1000}$ in Table 4), including ID and $L$ as input variables reduced (increased) the RMSE ($R^2$) value by 2.4% (1.0%). Comparison of the RMSE and $R^2$ values of SWRC1 with those of SWRC2 implies that including measurement scale parameters e.g., sample diameter and length *did not* improve the accuracy and reliability of pedotransfer functions. This means the MLR approach is not capable to detect interactions between input variables that define different subgroup of data with highly distinct predictor-response relationships, and to extract nonlinear patterns among input and output variables, in contrast to the CPXR technique (as shown in Table 3). In support, the values of RMSE given in Table 3 (resulted from CPXR) are 45-74% less than those presented in Table 4 (resulted from MLR) for both SWRC1 and SWRC2 models and the train and test processes.

Figure 2 shows the predicted water content versus measured one using the SWRC1 and SWRC2 models developed by the MLR method. The RMSE value of the SWRC2 model is slightly less than that of the SWRC1 model. The calculated RMSE values (0.0467 and 0.0416) of point pedotransfer functions developed by MLR presented in Fig. 2 are almost two times greater than those (0.0223 and 0.0226) of pedotransfer functions derived by CPXR reported in Fig. 1. Comparison of Figs. (1) and (2) also indicates that water content values predicted by MLR are more scattered than those predicted by CPXR, demonstrating the higher reliability of the proposed method in this study.

**4.2. Parametric pedotransfer functions for soil water retention curve (SWRC)**

The RMSE, RMSLE, and and $R^2$ values for the two SWRC3 and SWRC4 models developed using the CPXR method are given in Table 5. Comparison of the RMSE and RMSLE values of SWRC3 and SWRC4 models indicates that including measurement scale parameters increased the accuracy and reliability of the developed pedotransfer functions for predicting $\theta_s$, $\theta_r$, $\log_e(\alpha)$ and $\log_e(n)$, in both train and test processes. Particularly, in the train (test) process, the RMSE and RMSLE values of the pedotransfer functions developed for $\theta_r$, $\log_e(\alpha)$, and $\log_e(n)$ parameters decreased by 20% (21%), 35% (35%), and 49% (50%), respectively (see Table 5). However, the RMSE value of those developed for $\theta_s$ reduced only by 11% (16%). The obtained results confirm the effect of measurement scale on the soil water retention curve prediction, since $\alpha$ and $n$ parameters describe the shape of the soil water retention curve.

Table 5 also presents the $R^2$ values for the two SWRC3 and SWRC4 pedotransfer functions developed by the CPXR method. Comparison of the SWRC3 and SWRC4 models demonstrate that the $R^2$ value of the pedotransfer function developed for $\log_e(n)$ parameter increased considerably from 0.86 to 0.96 (12% increase) and 0.81 to 0.93 (15% increase) in the train and test processes, respectively. This means that the accurate prediction of the shape of the soil water retention curve requires information of the measurement scale. The $R^2$ values of the SWRC3 and SWRC4 models developed by CPXR presented in Table 5 are remarkably greater than those reported by Wösten et al.

(1995), Rajkai et al. (2004), and Merdun et al. (2006) for predicting van Genuchten soil water retention curve model parameters from soil textural data, organic matter, etc., which supports the application of the CPXR method and inclusion of measurement scale parameters in pedotransfer functions development.

Comparison of the RMSE values presented in Table 5 with those reported by Twarakavi et al. (2009) in their Table 3 indicates that the parametric pedotransfer functions developed by the CPXR method in this study predicted $\theta_s$ and $\theta_r$ more accurately than their models developed by the supporting vector machine method.

In Fig. 3, we present the predicted van Genuchten soil water retention curve model parameters versus fitted ones for the SWRC3 and SWRC4 models developed by the CPXR method. Comparison of the RMSE and RMSLE values of the SWRC4 model with those of the SWRC3 model implies that the SWRC4 model including the sample internal diameter and length predicted $\theta_s$, $\theta_r$, $\log_e(n)$, and $\log_e(\alpha)$ are more accurate than the SWRC3 model.

Figure 4 compares the predicted soil water retention curve using the parametric SWRC3 and SWRC4 models developed by the CPXR method with the measured one for the sample 4001 from the UNSODA database. As it demonstrates, the SWRC4 model in which measurement scale parameters e.g., sample internal diameter (ID) and length ($L$) were included as input parameters predicts the soil water retention curve more precisely than the SWRC3 model. Figure 4 also indicates that the SWRC3 model overestimated the soil water retention curve, while the SWRC4 model predictions match the measured values accurately.

The effect of column height on tension head and soil water retention curve has been addressed by Dane et al. (1992), Liu and Dane (1995), Perfect et al. (2004), among others. Since the density of nonwetting fluid (air) is different from the density of wetting fluid (water), in a tall column one should expect tension head varies with height (Dane et al., 1992; Liu and Dane, 1995), and thus one tension head value may not be representative for the entire column. The variation of tension head with height in a sample, however, would be negligible, if sample height is less than 2 cm (Dane and Hopmans, 2002). Regarding the effect of sample length on soil water retention curve model parameters, Perfect et al. (2004) corrected the measured soil water retention curves in lab using the method proposed by Liu and Dane (1995). Perfect et al. (2004) demonstrated that the Campbell's model (Campbell, 1974) parameters, such as pore-size distribution index and air entry value could be different with and without correction for column height.

The MLR results of the train and test processes are also given in Table 5. We found that the accuracy (RMSE) of the pedotransfer functions developed using the MLR method only improved by 5% and 4% in the train and test processes, respectively, when measurement scale parameters were included. Our results imply that the MLR method failed to detect patterns and construct nonlinear connections between input and output variables properly. However, the CPXR approach found effectively the nonlinear structures between inputs and outputs (see Table 5).

Figure 5 shows the predicted van Genuchten soil water retention curve model parameters as a function of fitted ones for the SWRC3 and SWRC4 models developed by MLR. Comparison of the RMSE values given in Fig 5 demonstrates that inclusion of the measurement scale parameters, e.g., sample internal diameter (ID) and length (*L*) not only does not improve the reliability of the developed parametric pedotransfer functions, but also may result in more uncertainties.

Comparison of Fig. 5 with Fig. 3 indicates remarkable difference in the accuracy of the pedotransfer functions developed by the MLR and CPXR methods. The RMSE values presented in Fig. 3 are considerably less than those reported in Fig. 5, which confirms the efficiency and robustness of the CPXR approach.

### 4.3. Pedotransfer functions for saturated hydraulic conductivity (SHC)

The accuracy and reliability of the pedotransfer functions (e.g., SHC1, SHC2, SHC3, and SHC4) developed to predict saturated hydraulic conductivity from other soil properties using the CPXR and MLR methods are summarized in Table 6.

Comparison of the RMSE values of the SCH1 and SHC2 models in the train process indicates that including sample internal diameter and length could decrease the RMSLE (root mean square logarithmic error) value from 0.964 to 0.547 (43% reduction) and 1.817 to 1.78 (2% reduction) using the CPXR and MLR methods, respectively. As can be seen, the RMSLE values of the SCH2 and SCH3 models are close meaning that the influence of measurement scale (ID and *L*) and soil water retention curve ($\theta_s$, $\theta_r$, $\alpha$ and *n*) parameters on the prediction of saturated hydraulic conductivity is comparable in both CPXR and MLR methods. However, comparison of the RMSLE value of the SCH4 model with that of other models indicated that inclusion of both measurement scale and soil water retention curve parameters increased model accuracy and reliability considerably (see Table 6). For example, in the CPXR method inclusion of measurement scale and soil water retention curve parameters (SHC4) decreased the RMSLE value by 59% (0.394 compared with 0.964 given in Table 6). However, including either measurement scale or soil water retention curve parameters only decreased the RMSLE value by about 44% (compare 0.547 with 0.964 or 0.532 with 0.964).

Comparison of the RMSLE values of the four SHC1, SHC2, SHC3, and SHC4 models given in Table 6 with those values reported in Table 5 of Twarakavi et al. (2009) showed that although the accuracy of the SHC1 model is less than their four different models, the SHC4 model developed by the CPXE method in this study predicted saturated hydraulic conductivity more accurately than the models developed by Twarakavi et al. (2009) using the supporting vector machine technique. We should point out that the accuracy of the SHC2 and SHC3 models is comparable and even more precise in some cases than those four models proposed by Twarakavi et al. (2009).

Figure 6 shows the predicted $\log(K_{sat})$ using the SHC1 to SHC4 models versus the measured one for one split of 10-fold cross-validation. As can be seen, the accuracy (RMSLE) of the developed models increased (decreased) when measurement scale and/or soil water retention curve parameters were included as input variables, in accordance with RMSLE values given in Table 6.

The results obtained in this study regarding the important effect of measurement scale on saturated hydraulic conductivity measurements and predictions are consistent with those reported in the literature. For example, Zobeck et al. (1985) studied the effect of sample cross-sectional area on saturated hydraulic conductivity in two structured clay soils. The soil cross-sectional areas utilized were 265 (soil blocks), 44 (soil cores), and 13 cm$^2$ (soil cores). For the first two sample sizes the constant head method, and for the latter the falling head permeameter procedure was used to measure SHC. Zobeck et al. (1985) found that in clay soils without macropores, all sample sizes produced similar mean SHC values, and recommended large soil blocks only in clay soils with well-developed structure and numerous macropores. The effect of sample volume on saturated hydraulic conductivity was investigated by Vepraskas and Williams (1995). They measured saturated hydraulic conductivity of quartz-diorite saprolite using sample volumes of 347, 6280, and 675000 cm$^3$, and demonstrated that the mean saturated hydraulic conductivity value found for samples with volumes of 347 cm$^3$ was significantly less than the mean values found for larger samples. They also indicated that the mean saturated hydraulic conductivity values of the sample volumes of 6280 and 675000 cm$^3$ were not significantly different. In another study, Fuentes and Flury (2005) investigated the effect of sample column length on the SHC of an undisturbed, no-till, silt loam soil. They showed that even small differences in core length could considerably affect the hydraulic conductivity measurement. For instance, the saturated hydraulic conductivity varied from 111 cm/day for the 15 cm sample length to 333 cm/day for the 25 cm sample length, with a coefficient of variation of 41%. More recently, Pachepsky et al. (2014) illustrated the similarity of scale dependences in soils and sediments (see Figs. 1 and 2 in Pachepsky et al., 2014). They indicated that as the characteristic support size increased, the SHC value first increased by one to two orders of magnitude and stabilized.

In the pedotransfer functions developed by the MLR method, inclusion of either measurement scale and/or soil water retention curve parameters decreased the RMSLE value only by 7% (see Table 6). This value compared to that obtained from the CPXR method (59%) implies that the CPXR technique provides a robust, effective algorithm to find nonlinear patterns between input and output variables.

In Fig. 7, we present the results of the developed saturated hydraulic conductivity pedotransfer functions (SHC1 to SHC4 models) using the MLR method. As the calculated RMSE values indicate including measurement scale and/or van Genuchten soil water retention curve model parameters increased the reliability of the models. Comparing Fig. 7 with Fig. 6 also demonstrates that the saturated hydraulic conductivity predictions of the MLR method are considerably more scattered than those of the CPXR approach. The calculated RMSLE values given in Figs. 6 and 7 also support that the CPXR-based pedotransfer functions predict the saturated hydraulic conductivity more accurately than the MLR-based ones.

## 5. Conclusion

In this study, pedotransfer functions were developed using contrast pattern aided regression (CPXR) and multiple linear regression (MLR) methods to predict the soil water retention curve and the saturated hydraulic conductivity. For this purpose, the 10-fold cross-validation approach was applied to evaluate the developed models accuracy and reliability. Two databases were selected from the UNSODA database including soil samples whose measurement scale parameters, e.g., sample internal diameter and length were available. The first database, consisting of 210 soils, was used to develop point and parametric pedotransfer functions for the soil water retention curve, and the second one, including 213 samples, for the saturated hydraulic conductivity prediction. The obtained results indicated that the CPXR method predicted output variables more accurately than the MLR technique in both train and test steps. As expected, we demonstrated that inclusion of sample internal diameter and length could improve the accuracy and reliability of the developed pedotransfer functions remarkably.

Table 1. An arbitrary dataset including four soil samples

| Sample # | Sand (%) | Silt (%) | Clay (%) |
|---|---|---|---|
| 1 | 83 | 14 | 3 |
| 2 | 84.6 | 14.4 | 3 |
| 3 | 81.5 | 15 | 2 |
| 4 | 85 | 13 | 1 |

Table 2. Input and output variables of different models developed in this study.

| Model | Input variables | Output variables |
| --- | --- | --- |
| SWRC1 (point) | Sa, Si, Cl, $\rho_b$, $d_g$, $\sigma_g$ | $\theta_s$, $\theta_i$, $\theta_{10}$, $\theta_{30}$, $\theta_{50}$, $\theta_{100}$, $\theta_{300}$, $\theta_{500}$, $\theta_{1000}$, $\theta_{1500}$ |
| SWRC2 (point) | Sa, Si, Cl, $\rho_b$, $d_g$, $\sigma_g$, ID, $L$ | $\theta_s$, $\theta_i$, $\theta_{10}$, $\theta_{30}$, $\theta_{50}$, $\theta_{100}$, $\theta_{300}$, $\theta_{500}$, $\theta_{1000}$, $\theta_{1500}$ |
| SWRC3 (parametric) | Sa, Si, Cl, $\rho_b$, $d_g$, $\sigma_g$ | $\theta_r$, $\theta_s$, $\log_e(\alpha)$, $\log_e(n)$ |
| SWRC4 (parametric) | Sa, Si, Cl, $\rho_b$, $d_g$, $\sigma_g$, ID, $L$ | $\theta_r$, $\theta_s$, $\log_e(\alpha)$, $\log_e(n)$ |
| SHC1 | Sa, Si, Cl, $\rho_b$, $d_g$, $\sigma_g$ | $\log_e(K_{sat})$ |
| SHC2 | Sa, Si, Cl, $\rho_b$, $d_g$, $\sigma_g$, ID, $L$ | $\log_e(K_{sat})$ |
| SHC3 | Sa, Si, Cl, $\rho_b$, $d_g$, $\sigma_g$, $\theta_r$, $\theta_s$, $\alpha$, $n$ | $\log_e(K_{sat})$ |
| SHC4 | Sa, Si, Cl, $\rho_b$, $d_g$, $\sigma_g$, $\theta_r$, $\theta_s$, $\alpha$, $n$, ID, $L$[*] | $\log_e(K_{sat})$ |

[*] Sa: sand, Si: silt, Cl: clay, $\rho_b$: bulk density, $d_g$: geometric mean diameter, $\sigma_g$: geometric standard deviation, $\theta_r$, $\theta_s$, $\alpha$, $n$: van Genuchten soil water retention curve model parameters, ID: sample internal diameter, $L$: sample length, SWRC: soil water retention curve, SHC: saturated hydraulic conductivity, and $\log_e$: natural logarithm.

Table 3. Statistical parameters (RMSE and $R^2$) calculated for the train and test splits and point pedotransfer functions of soil water retention curve (SWRC1 and SWRC2) using the CPXR method.

|  | Model | $\theta_s$ | $\theta_i$ | $\theta_{10}$ | $\theta_{30}$ | $\theta_{50}$ | $\theta_{100}$ | $\theta_{300}$ | $\theta_{500}$ | $\theta_{1000}$ | $\theta_{1500}$ |
|---|---|---|---|---|---|---|---|---|---|---|---|
|  |  |  |  |  |  | RMSE |  |  |  |  |  |
| Train | SWRC1 | 0.018 | 0.013 | 0.030 | 0.025 | 0.023 | 0.020 | 0.017 | 0.017 | 0.019 | 0.021 |
|  | SWRC2 | 0.016 | 0.011 | 0.018 | 0.018 | 0.018 | 0.017 | 0.014 | 0.013 | 0.014 | 0.016 |
| Test | SWRC1 | 0.019 | 0.013 | 0.029 | 0.026 | 0.023 | 0.021 | 0.018 | 0.019 | 0.020 | 0.025 |
|  | SWRC2 | 0.016 | 0.012 | 0.020 | 0.019 | 0.019 | 0.022 | 0.020 | 0.016 | 0.015 | 0.017 |
|  |  |  |  |  |  | $R^2$ |  |  |  |  |  |
| Train | SWRC1 | 0.96 | 0.98 | 0.96 | 0.96 | 0.97 | 0.97 | 0.97 | 0.97 | 0.96 | 0.95 |
|  | SWRC2 | 0.97 | 0.98 | 0.98 | 0.98 | 0.98 | 0.98 | 0.98 | 0.98 | 0.98 | 0.97 |
| Test | SWRC1 | 0.94 | 0.97 | 0.97 | 0.94 | 0.97 | 0.97 | 0.95 | 0.96 | 0.95 | 0.94 |
|  | SWRC2 | 0.95 | 0.96 | 0.94 | 0.95 | 0.97 | 0.96 | 0.95 | 0.98 | 0.97 | 0.94 |

Table 4. Statistical parameters (RMSE and $R^2$) calculated for the train and test splits and point pedotransfer functions of soil water retention curve (SWRC1 and SWRC2) using the MLR method.

|  | Model | $\theta_s$ | $\theta_i$ | $\theta_{10}$ | $\theta_{30}$ | $\theta_{50}$ | $\theta_{100}$ | $\theta_{300}$ | $\theta_{500}$ | $\theta_{1000}$ | $\theta_{1500}$ |
|---|---|---|---|---|---|---|---|---|---|---|---|
|  |  |  |  |  |  | RMSE |  |  |  |  |  |
| Train | SWRC1 | 0.062 | 0.042 | 0.062 | 0.048 | 0.045 | 0.041 | 0.039 | 0.039 | 0.041 | 0.045 |
|  | SWRC2 | 0.061 | 0.042 | 0.061 | 0.048 | 0.045 | 0.041 | 0.039 | 0.039 | 0.040 | 0.045 |
| Test | SWRC1 | 0.062 | 0.042 | 0.062 | 0.048 | 0.045 | 0.041 | 0.039 | 0.040 | 0.041 | 0.045 |
|  | SWRC2 | 0.061 | 0.042 | 0.062 | 0.049 | 0.045 | 0.041 | 0.040 | 0.039 | 0.041 | 0.045 |
|  |  |  |  |  |  | $R^2$ |  |  |  |  |  |
| Train | SWRC1 | 0.81 | 0.76 | 0.81 | 0.87 | 0.88 | 0.88 | 0.87 | 0.86 | 0.84 | 0.80 |
|  | SWRC2 | 0.82 | 0.76 | 0.81 | 0.87 | 0.88 | 0.88 | 0.87 | 0.86 | 0.84 | 0.80 |
| Test | SWRC1 | 0.79 | 0.73 | 0.77 | 0.84 | 0.85 | 0.84 | 0.83 | 0.84 | 0.81 | 0.77 |
|  | SWRC2 | 0.79 | 0.72 | 0.77 | 0.85 | 0.84 | 0.84 | 0.84 | 0.83 | 0.80 | 0.78 |

Table 5. Statistical parameters (RMSE, RMSLE and $R^2$) calculated for the train and test splits and parametric pedotransfer functions of the van Genuchten soil water retention curve model (SWRC3 and SWRC4) using the CPXR and MLR approaches.

| Method | | Model | $\theta_r$ | $\theta_s$ | $\log_e(\alpha)$ | $\log_e(n)$ | $\theta_r$ | $\theta_s$ | $\log_e(\alpha)$ | $\log_e(n)$ |
|---|---|---|---|---|---|---|---|---|---|---|
| | | | RMSE | | RMSLE* | | $R^2$ | | | |
| CPXR | Train | SWRC3 | 0.030 | 0.018 | 0.531 | 0.191 | 0.80 | 0.96 | 0.83 | 0.86 |
| | | SWRC4 | 0.024 | 0.016 | 0.346 | 0.098 | 0.87 | 0.97 | 0.93 | 0.96 |
| | Test | SWRC3 | 0.034 | 0.019 | 0.570 | 0.201 | 0.76 | 0.94 | 0.83 | 0.81 |
| | | SWRC4 | 0.027 | 0.016 | 0.371 | 0.101 | 0.83 | 0.95 | 0.90 | 0.93 |
| MLR | Train | SWRC3 | 0.060 | 0.055 | 1.136 | 0.328 | 0.19 | 0.60 | 0.21 | 0.69 |
| | | SWRC4 | 0.060 | 0.054 | 1.158 | 0.319 | 0.19 | 0.62 | 0.18 | 0.61 |
| | Test | SWRC3 | 0.061 | 0.055 | 1.140 | 0.330 | 0.21 | 0.58 | 0.20 | 0.58 |
| | | SWRC4 | 0.060 | 0.053 | 1.167 | 0.324 | 0.22 | 0.59 | 0.18 | 0.59 |

*RMSLE is root mean square logarithmic error

Table 6. Statistical parameters (RMSLE and $R^2$) calculated for the train and test splits and saturated hydraulic conductivity pedotransfer functions (SHC1, SHC2, SHC3, and SHC4) using CPXR and MLR method.

| Method | | Train | | | | Test | | | |
|---|---|---|---|---|---|---|---|---|---|
| | | SHC1 | SHC2 | SHC3 | SHC4 | SHC1 | SHC2 | SHC3 | SHC4 |
| CPXR | RMSLE* | 0.964 | 0.547 | 0.532 | 0.394 | 0.987 | 0.57 | 0.551 | 0.421 |
| | $R^2$ | 0.83 | 0.94 | 0.95 | 0.97 | 0.82 | 0.91 | 0.93 | 0.94 |
| MLR | RMSLE | 1.817 | 1.780 | 1.730 | 1.695 | 2.1 | 1.805 | 1.843 | 0.712 |
| | $R^2$ | 0.39 | 0.42 | 0.45 | 0.47 | 0.28 | 0.35 | 0.38 | 0.41 |

* RMSLE is root mean square logarithmic error

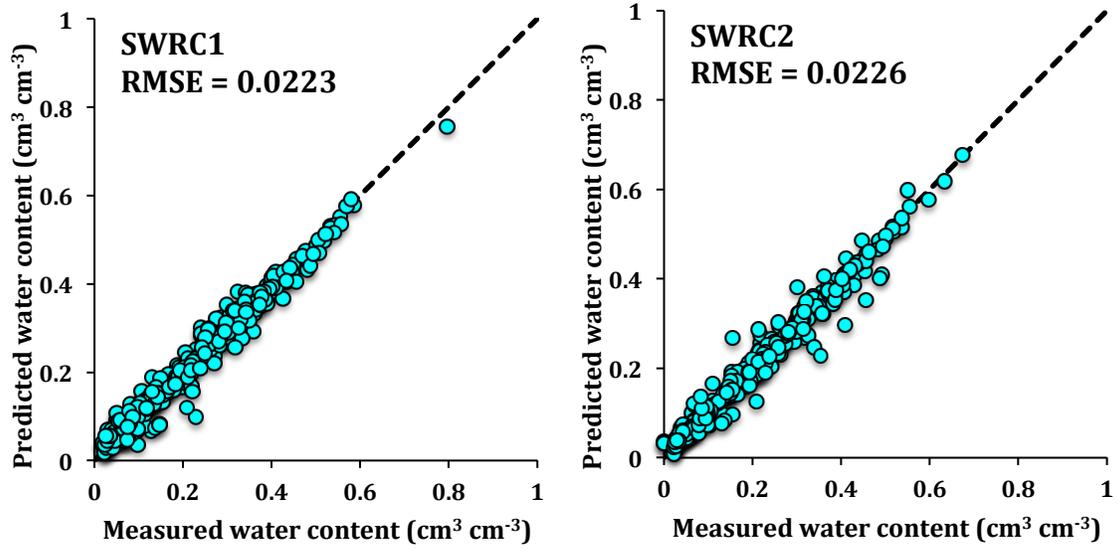

Fig. 1. Predicted water content versus measured one using the CPXR method for SWRC1 and SWRC2 point pedotransfer functions developed in this study. Results represent one split of 10 repetitions of 10-fold cross-validations.

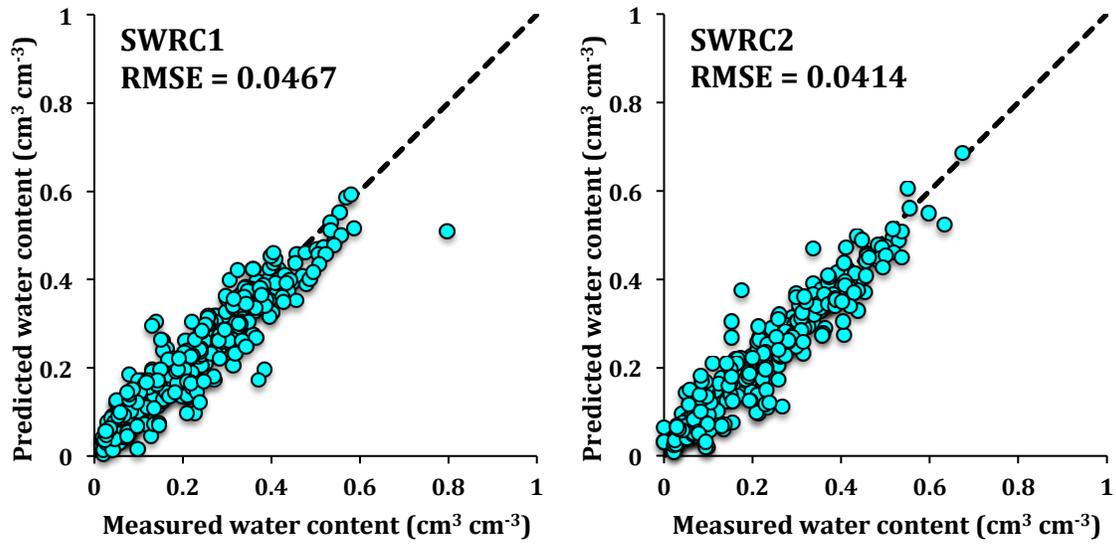

Fig. 2. Predicted water content versus measured one using the MLR method for SWRC1 and SWRC2 point pedotransfer functions developed in this study. Results represent one split of 10 repetitions of 10-fold cross-validations.

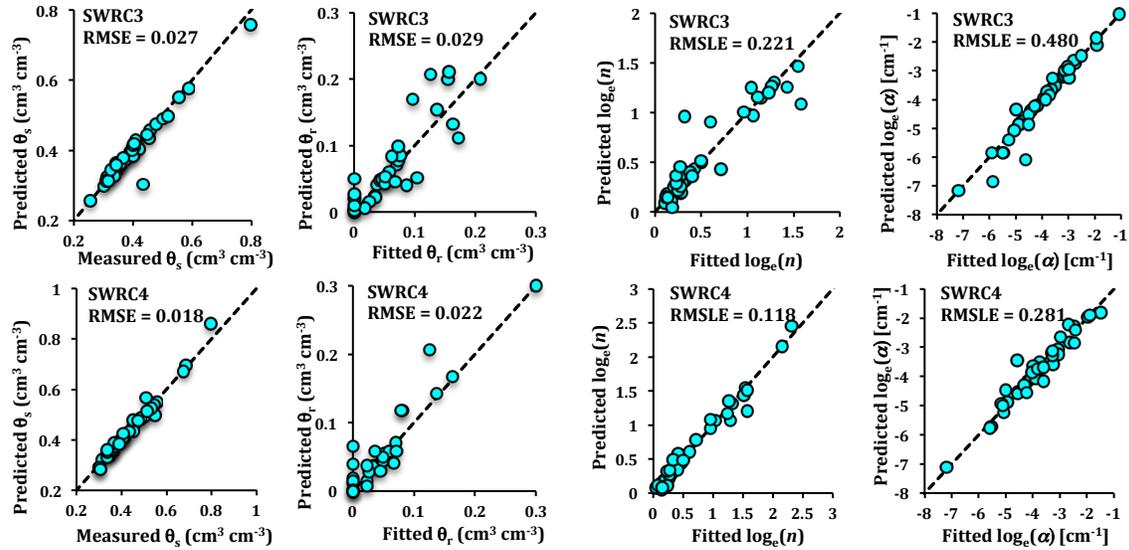

Fig. 3. Predicted van Genuchten water retention curve model parameters versus fitted ones using the CPXR method for SWRC3 and SWRC4 parametric pedotransfer functions developed in this study. Results of each plot represent one split of 10 repetitions of 10-fold cross-validations.

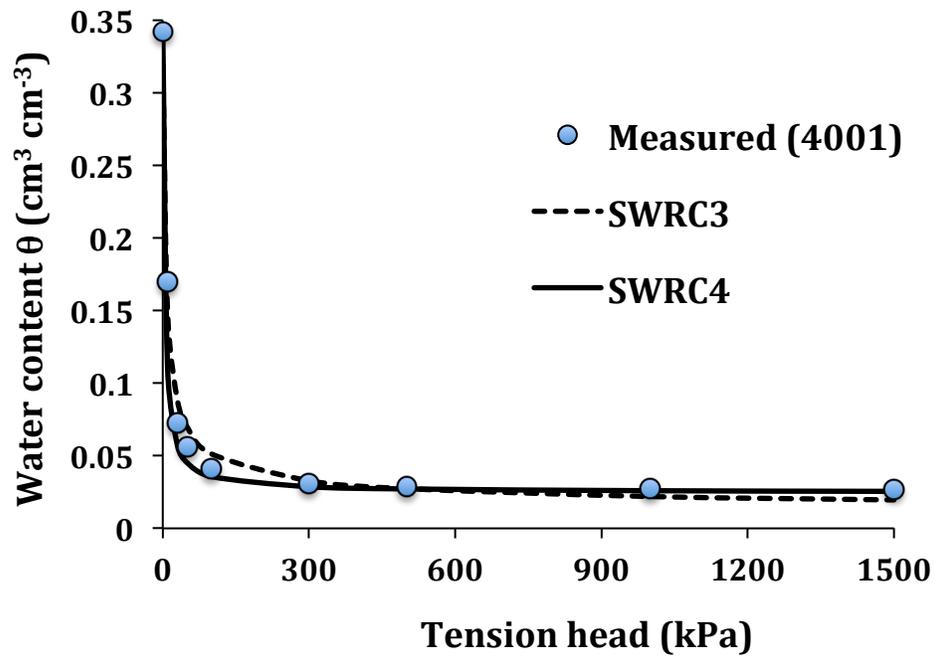

Fig. 4. Comparison of the predicted soil water retention curve using parametric SWRC3 and SWRC4 models developed by the CPXR method in this study with the measured one for sample 4001 from the UNSODA database. This figure shows that including sample internal diameter and length improves soil water retention curve noticeably. Note that the results represent one split of 10 repetitions of 10-fold cross-validations.

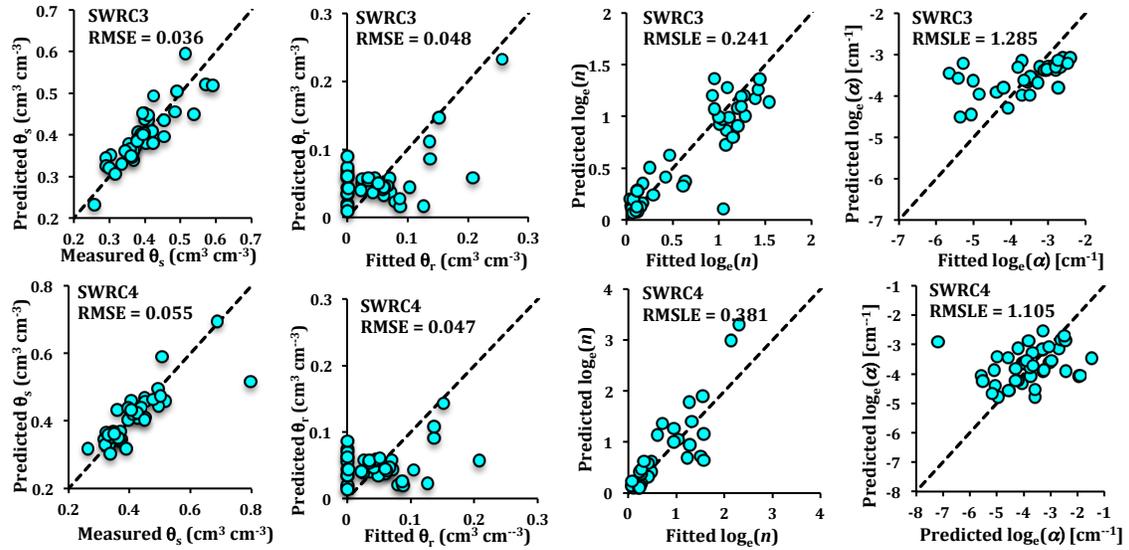

Fig. 5. Predicted van Genuchten water retention curve model parameters versus fitted ones using the MLR method for SWRC3 and SWRC4 parametric pedotransfer functions developed in this study. Results of each plot represent one split of 10 repetitions of 10-fold cross-validations.

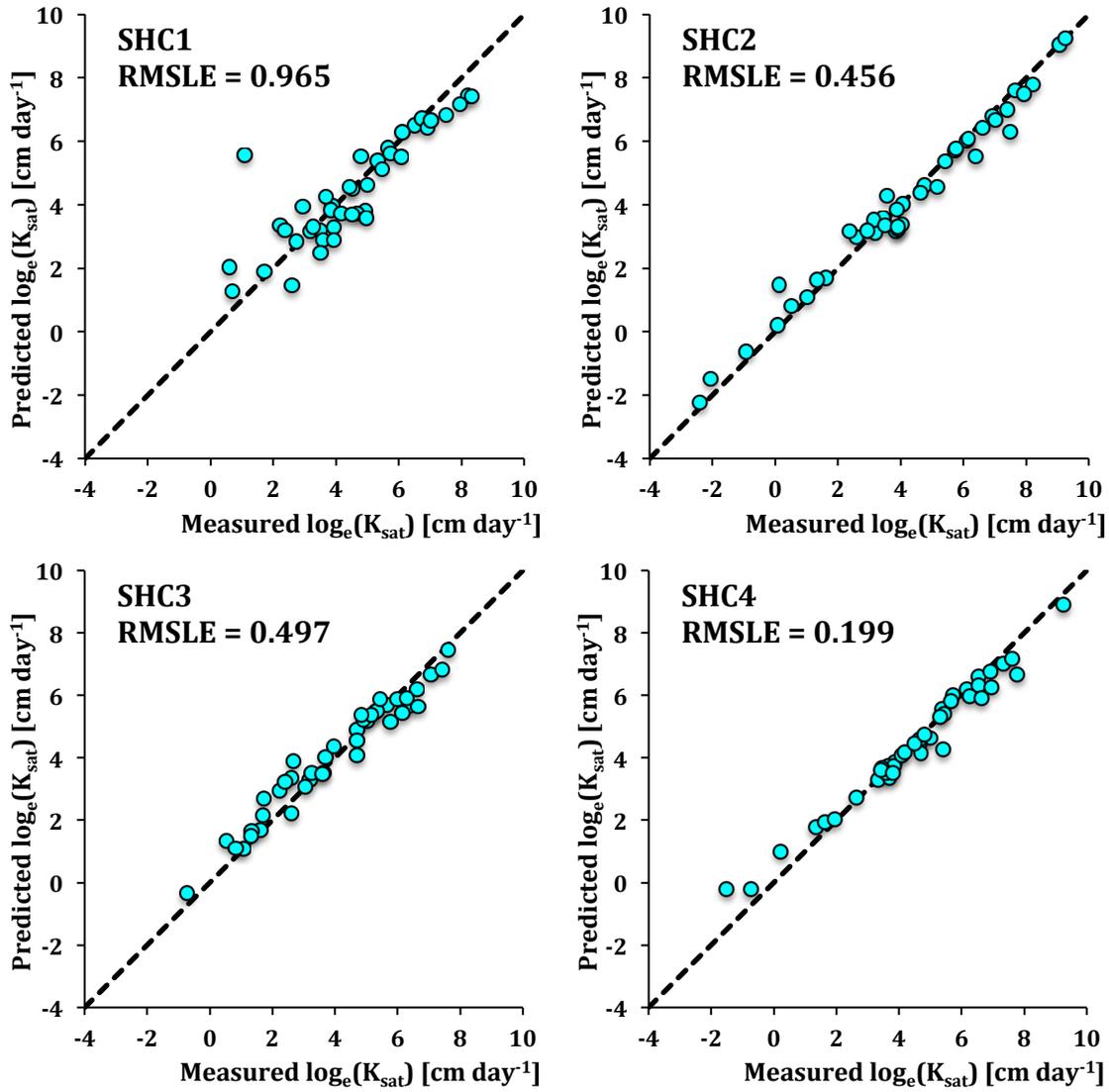

Fig. 6. Predicted saturated hydraulic conductivity versus measured one for 4 models with different input variables developed using the CPXR method. Results represent one split of 10 repetitions of 10-fold cross-validations, and RMSLE is root mean square logarithmic error.

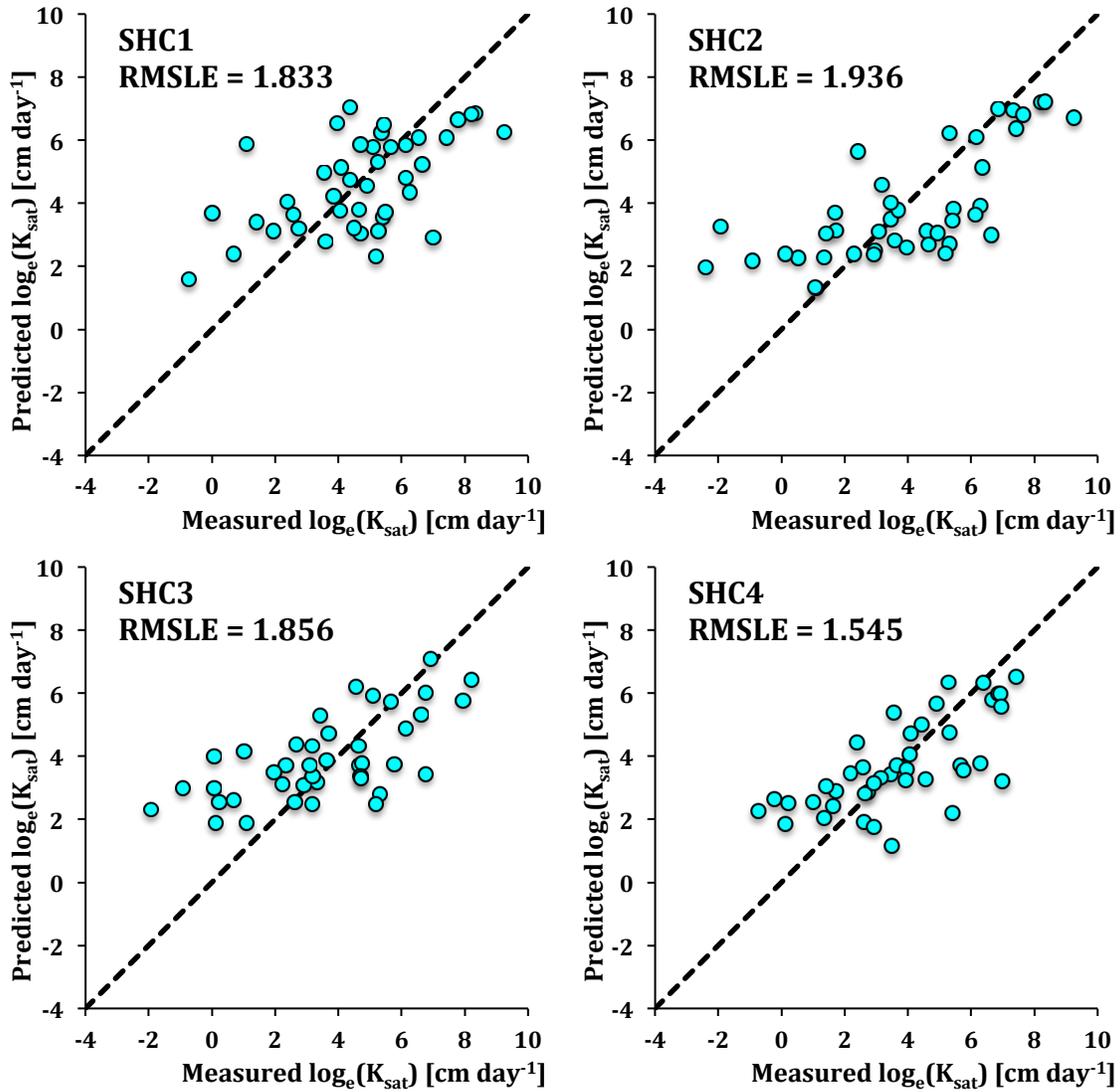

Fig. 7. Predicted saturated hydraulic conductivity versus measured one for 4 models with different input variables developed using the MLR method. Results represent one split of 10 repetitions of 10-fold cross-validations, and RMSLE is root mean square logarithmic error.